\documentclass[12pt,aps]{revtex4}
\usepackage{epsfig}
\usepackage{amsmath,amssymb,color}
\usepackage{epstopdf}
\usepackage{braket}
\usepackage{graphicx}
\usepackage{hyperref}
\hypersetup{
    colorlinks=true,
    linkcolor=cyan,
    filecolor=blue,      
    citecolor=blue
}
\usepackage[draft]{todonotes} 
\usepackage{multirow}
\parskip=\medskipamount

\newcommand{\beq}{\begin{equation}}
\newcommand{\eeq}{\end{equation}}
\newcommand{\be}{\begin{equation}}
\newcommand{\ee}{\end{equation}}
\newcommand{\bal}{\begin{align}}
\newcommand{\eal}{\end{align}}

\newcommand{\la}{\left<}

\def \la {\lambda}

\begin{document}

\title{Metal or Insulator? Dirac operator spectrum 
in holographic QCD}

\author{A. Gorsky$^{2,3}$, M. Litvinov$^{1,3}$}

\affiliation{
$^1$ Skolkovo Institute of Science and Technology, Moscow, Russia 121205\\
$^2$ Institute for Information Transmission Problems of the Russian Academy of Sciences, Moscow,
Russia 127051\\
$^3$ Moscow Institute of Physics and Technology, Dolgoprudny 141700, Russia \\
}

\begin{abstract}
The lattice studies in QCD demonstrate the nontrivial localization behavior
of the eigenmodes of the 4D Euclidean Dirac operator considered as Hamiltonian 
of $4+1$ dimensional disordered system. We use the holographic viewpoint to provide
the conjectural explanation of these properties. 
The delocalization of all modes in the confined phase is related to the
$\theta=\pi$ - like phenomena when the fermions are delocalized on domain walls. 
It is conjectured that 
the localized modes separated by mobility edge from the rest of the spectrum 
in deconfined QCD correspond to the near-horizon region
in the holographic dual.

\end{abstract}
\maketitle

\section{Introduction } 

The 4D Euclidean Dirac operator  $i\gamma_\mu D^{\mu}$ spectrum in QCD is the important
observable both in the confined and deconfined phases.
For instance, the  Casher-Banks relation \cite{casher}  relating the chiral condensate 
with the spectral density reads as
\begin{align}
\braket{\overline{\psi}\psi}=\lim_{\lambda\to0}\lim_{m\to0}\frac{\pi\rho(\lambda,m)}{V},
 \end{align}
where $m$ is the current quark mass and $V$ is the four-volume. Two  matrix  models  are  useful  for  a  investigation  of the Dirac operator spectral properties in confined phase (see \cite{matrix} for review).  One of them corresponds to  zero momentum sector of the Chiral Lagrangian while the second model  mimics  the  fermion  determinant  integrated  over  the  moduli  space  of  a  instanton-antiinstanton ensemble presumably relevant for QCD ground state.

The spectral problem for the
Dirac operator  has been treated in \cite{stern} by the tools familiar in a theory of  disordered systems. 
The Euclidean 
Dirac operator in 4D was considered as the disordered Hamiltonian providing evolution in  the additional
fifth time coordinate identified as the Schwinger proper time \cite{stern}. On the other 
hand the proper time is related with  the radial coordinate in the ADS-like geometry
in Poincare metric \cite{gopakumar,gl}. This is the starting point for our analysis of the holographic
treatment of the Dirac operator spectrum. Since the radial evolution can be identified
with the RG flow \cite{deboer} our consideration to some extend deals with specific aspects of the RG flows
in QCD.

We consider the disordered (4+1) Hamiltonian hence the immediate  standard question concerning the localization properties of the eigenmodes of the Dirac operator in Euclidean 4D space arises. 
The delocalized 
modes are subject to the level interaction and obey the Wigner-Dyson
statistics, while  the localized modes do not interact at all and obey the
Poisson statistics. There can be a mobility edge separating  
localized and delocalized modes in the $d\geq 3$, where $d$ is dimension
of space. In $1+1$ and $2+1$  dimensions the most of modes are localized, however,
there could be a few delocalized modes if the topological terms
are present in the action \cite{khmel,pruisken,furusaki,kamenev}. For instance, 
these distinguished delocalized modes are responsible for the Hall
conductivity in $2+1$ case.

The properties of the Dirac operator spectrum 
have been investigated in the lattice QCD and the results found
were a bit surprising. They can be summarized as follows:
\begin{itemize}

\item  All modes are delocalized in the confined phase \cite{osborn}

\item There is the mobility edge $\lambda_m$ in the deconfined phase  \cite{osborn,kovacs}. Low energy
modes in the deconfined phase are localized while high energy part of the
spectrum is delocalized.

\item 
The  mobility edge $\lambda_m(T)$
at $T>T_c$ grows as the function of the temperature near the deconfinement 
phase transition approximately as \cite{temperature}
$
\lambda_m(T) = a (T-T_c)
$
with come constant $a$. The fractal dimension at the localization phase transition coincides
with the   fractal dimension of 3D unitary Anderson model \cite{uam}.

\end{itemize}
The brief summary of these results can be found in \cite{giordano}.

In this Letter we consider the localization properties of the Dirac operator spectrum from the
holographic viewpoint. 
The deconfinement transition  holographically corresponds to the change of the bulk geometry which involves
the thermal AdS at $T<T_c$ and AdS-like black hole (BH) at $T>T_c$ \cite{witten98,aharonydec}. 
We assume that the delocalization of all modes in the confined phase is related to the 
existence of the domain walls at $\theta=\pi$. The relevance of this regime follows
from the fact that the eigenfunction of the Dirac operator corresponds to the 
quark with imaginary mass and the phase of the mass is traded by the axial anomaly to the
non-vanishing $\theta$-term. The quarks are deconfined at the domain wall \cite{zohar}
hence this interpretation implies that the quark propagation in 5-th time occurs along 
the $2+1$- dimensional subspace of 4d Euclidean space. This qualitatively fits
the lattice results \cite{qcdOLD}. 
It is also natural 
to question how the emergence of BH in the deconfined phase and 
the mobility edge in the Euclidean  4D Dirac operator spectrum are correlated. We
have found the evidences that the mobility edge corresponds to the near-horizon region.

It is worth to make one more comment.
It was argued in \cite{fradkin} that the disorder driven transition for the
Dirac operator differs from the Anderson transition for the non-relativistic  Schrodinger
operator. The key difference
concerns the role of spectral density  as the order parameter.
In the usual Anderson transition  the spectral density does not play
any essential role and only a spectral formfactor 
and higher spectral correlators matter. We shall be interested in the Dirac operator spectrum in $4D$
Euclidean space hence the non-Anderson transition can be expected. 

The Letter is organized as follows. In Section 2 we collect
relevant properties concerning the Dirac operator in QCD. In Section 3 we briefly review the holographic model of QCD. In Section 4 we 
consider the different approaches familiar in the theory of disordered systems
to diagnose the localization properties of  Hamiltonian with
disorder of different nature. In Section 5 we relate
the delocalization of all modes in the confined phase  with 
the $\theta=\pi$ - like phenomena and assume that the  quarks propagate
along the domain wall where they are deconfined. In Section 6
we conjecture that the  localized modes separated by mobility edge from the rest of the spectrum 
in the deconfined phase 
correspond to the near-horizon region in the holographically
dual  black hole. Some open questions are formulated in the Discussion.

\section{Dirac operator in QCD}

Let us recall some results concerning the Dirac operator spectrum.
The partition function of QCD reads as 
\beq
Z_{QCD}= \int dA_{\mu}\prod_{f=1}^{N_f} \det(iD+m_f)\exp(-S_{YM}(A)).
\eeq
We shall deal with the eigenvalue equation  for the 4D Euclidean Dirac operator
\beq
\hat{D}(A)\psi_n=i\lambda_n\psi_n,
\eeq
which coincides with the Dirac equation for the imaginary fermion 
mass $m=i\lambda$.
The spectral density is defined as 
\beq
\rho(\lambda)=\Braket{\sum_n \delta(\lambda- \lambda_n)}_{QCD}
\eeq
and near the origin in the confined phase it behaves as  \cite{casher,smilga}
\beq
\rho(\la) =\frac{\pi \rho(0)}{V} + c |\lambda|+ O(\lambda)
\label{density}
\eeq
where \cite{smilga}
\beq
c=\frac{\rho(0)^2(N_f^2 -4)}{32N_fF_{\pi}^2}.
\eeq
The spectral density can be derived from the discontinuity of the resolvent
across the imaginary axis
\beq
\Sigma(z)=\frac{1}{V}\Braket{\operatorname{Tr}\frac{1}{D+z}},\text{  where $V$ is a volume.}\eeq

\section{Holographic preliminaries}
 Turn now to the holographic QCD and consider the Witten-Sakai-Sugimoto geometry \cite{witten98,ss}
\begin{eqnarray}
\begin{array}{ccccccccccc}
& 0 & 1 & 2 & 3 & (4) & 5 & 6 & 7 & 8 & 9 \\
\mbox{D4} & \circ & \circ & \circ & \circ & \circ &&&&& \\
\mbox{D8-$\overline{\text{D8}}$}
& \circ & \circ & \circ & \circ &  & \circ & \circ & \circ &\circ & \circ 

\end{array}
\label{D4D8}
\end{eqnarray}
It involves the $N_c$ D4 branes wrapped around the cylinder with the boundary
conditions for fermions breaking SUSY. At large $N_c$ the D4 branes pinch the cylinder which
turns into a cigar. The total 10D geometry
looks as $R^{3,1}\times S^4 \times (r,\phi)$ at small temperature.
In the confined phase  
the cigar in $(r,\phi)$ coordinates reads as
\beq
ds^2=(r/R)^{3/2}f(r)d\phi^2 + (R/r)^{3/2}\frac{dr^2}{f(r)} \qquad f(r)= 1 -\left(\frac{r_{kk}}{r}\right)^3 
\eeq 
where $\phi$ is periodic variable.
We insert D8-branes extended 
in radial coordinate r and localized at $\phi$ and D0 instantons
localized in radial coordinate in this background geometry
and extended along $\phi$. 
The $N_f$  D8$-\overline{\text{D8}}$ 
branes are connected at the tip of the cigar and are placed at 
$\phi=0,\pi$ on the $\phi$ circle.
The D8 branes carry $U(N_F)$ flavor gauge group
at the worldvolume and matrix $U$ of the pseudoscalar mesons $\pi_a$ is defined in terms of 
holonomy of radial component of the flavor
gauge field
\beq
U= e^{it_a\pi_a}= e^{\int A_rdr}
\eeq
Above the critical temperature the metric involves BH \cite{witten98}
and the phase transition in QCD qualitatively corresponds to the Hawking-Page transition.
At $T=T_c$ the $\phi$ and Euclidean time $t_E$ 
coordinates get interchanged.

It is worth also to comment on the holographic origin of the mass term $\operatorname{Tr}MU$  in the Chiral Lagrangian.
It was argued in \cite{aharony2,hashimoto} that it  comes from the 
worldsheet instanton that is the open string with worldsheet $(r,\phi)$ coordinates
which is stretched between left and right D8 branes and spans some area on the cigar.
The mass comes from the Nambu-Goto string action while the factor $U$ comes 
from the interaction of the open string end with D8 brane. We are interested 
in the Dirac operator eigenvalues that is purely imaginary masses. The imaginary masses can be
obtained if the $\theta$ term is taken into account which holographically corresponds
to the holonomy of the RR 1-form field along KK circle \cite{wittentheta}. With the proper
value of $\theta$ we get purely imaginary masses. The early discussion on the Dirac
operator spectral density in holographic QCD can be found in \cite{kopnin}.

\section{Critical regime}
\subsection{Diagnostics of the critical behavior}

In what follows we shall be interested in the spectral properties of 
the disordered Hamiltonian
near the mobility edge $E_m$. 
There are several specific features intrinsic for this regime
supporting the multifractal behavior.
\begin{itemize}
\item

First, the level spacing distribution $P(s)$ is the key indicator
of the localization/delocalization transition. It behaves as 
\beq
\left\{
\begin{array}{ll}
P_{deloc}(s)= A\, s^2\,e^{-Bs^2} & \mbox{delocalized phase(GOE)}
\medskip \\ P_{loc}(s) = e^{-\frac{s}{2\chi}} & \mbox{localized phase}
\end{array} \right.
\label{eq:05}
\eeq
where for unfolded spectrum  $s_i= x_{i+1} - x_i$, $ x_i = \int^{\lambda_i} \rho(\lambda)$. The mean level spacing is $\displaystyle \Delta\equiv\frac{1}{\braket{\rho(0)}}$.
Let us take a window of the width $\delta E$, $\displaystyle\frac{\delta E}{\Delta} \equiv \overline{n}  \ll N$,
in the energy space centered at $E = 0$ and calculate the number $n$ of levels inside the window
at some realization of disorder.  The parameter $\chi$ in the Poisson tail is  the level compressibility defined as 
\beq
\chi = \frac{d}{d\bar{n}}\Braket{(n-\bar{n})^2}, \qquad N\gg \bar{n}\gg 1
\eeq

\item
The second indicator is the  spectral correlator which develops a fractal behavior at the mobility
edge which differs both from Wigner-Dyson and Poisson statistics \cite{edge}
\beq
R(E)\propto E ^{-1+\frac{D_2}{d}}
\eeq
where $D_2$ is the fractal dimension defined as 
\beq
\sum_{r,n}\Braket{|\Psi_n(r)|^{2p}\delta(E-E_n)}\propto L^{-D_p(p-1)}
\eeq
in the volume $L^d$, $d$ is dimension of space. 
The level number variance 
\beq
\Sigma =\Braket{(n-\bar{n})^2}
\eeq
behaves  as  $\Sigma_{crit} \propto \chi E$ 
in the multifractal case, where the level compressibility reads as
\beq
\chi =  \frac{d-D_2}{2d}
\label{level}
\eeq

\end{itemize}

\subsection{Matrix models for localization transition}

There are several critical matrix 
models \cite{crit1,crit2,crit3} describing
the  localization transition in 3D 
which are qualitatively unified in \cite{muttalib}. All of them 
works only nearby the mobility edge
(see \cite{kravtsovrev} for review). Let us summarize their main features 

\begin{itemize}
\item
The two-matrix model \cite{crit1} involves the
following probability function 
\beq
P(H) \propto \exp\left(-\beta \operatorname{Tr} H^2 - \beta b \operatorname{Tr}([\Omega,H][\Omega,H]^{\dagger})\right)
\eeq
where $b$ is parameter and $\Omega$ is the fixed unitary matrix $\displaystyle \Omega = \operatorname{diag}\left(\exp\left(\frac{2\pi i k}{N}\right)\right)$.
The critical regime in this model implies that unitary symmetry breaking parameter 
behaves as $b=\mu N^2$ at $N\rightarrow \infty$.

\item

The second  one-matrix model  involves the potential providing
the weak confinement \cite{crit2} of eigenvalues. 
Asymptotically potential behaves as 
\beq
V(x)\rightarrow \frac{1}{\gamma}\log^2 x, \qquad x\rightarrow \infty,
\eeq
where $\gamma$ - is some parameter and the probability measure
in the matrix integral reads as 
\beq
P(H)\propto \exp \big(-\beta \operatorname{Tr}V(H) \big)
\eeq
The similar critical model for the  chiral ensembles has been 
considered in \cite{vercrit}.

\item

The third model was suggested in \cite{crit3} and involves the Gaussian
ensemble of independent random entries $(i\geq j)$
\beq
\Braket{H_{ij}}=0,\qquad \Braket{(H_{ij})^2}=\beta^{-1}\left[1 +\frac{(i-j)^2}{B^2}\right]^{-1}
\eeq
where $B$ is parameter of the model, for $B>>1$ it gets mapped into supersymmetric sigma model. This 
model also manifests the multifractal behavior \cite{crit3} at the mobility edge.

\end{itemize}

The  spectral correlators in all three
critical models are the same
\beq
R(E,t)=\Braket{\rho(E)\rho(E+t)}= \delta(t) + Y(E,t)
\eeq
where at small $t$
\beq
Y(E,t)\propto \frac{\pi^2\eta^2}{4} \frac{\sin^2(\pi t)}{\sinh^2(\pi^2 t \eta/2)}
\qquad \beta=2
\eeq
The parameter $\eta$ is related with the parameters of the matrix ensembles
$\displaystyle\eta= \sqrt{\mu}=\frac{\gamma}{\pi^2} $ if we assume $\eta << 1$. The spectral correlator 
in this regime is identical to 
the density-density correlator for the free fermion gas at finite temperature
proportional to $\eta$. The parameter $\eta $ is related with the fractal
dimension as $\eta = 1 - D_2$. Similarly at small $\eta$ regime 
the spectral compressibility reads as
\beq
\chi=1 + \int_{-\infty}^{+\infty} Y(E,t)dt
\eeq
and is consistent with the general relation (\ref{level}).

Let us remark that formulas above are valid only for small
multifractality. At large $t$ there is the power-tail
which knows about the fractal dimension as well.
According to our conjecture the small $t$ regime
corresponds to the near  horizon IR region  while
the large $t$ regime captures the information about
the UV scale. Since we are dealing with a kind of anomaly
phenomena the information about the fractal dimension
can be captured both in UV and IR regions.

\section{Spectral statistics in the confined phase}
\subsection{ Towards the mechanism for delocalization}

The lattice studies  demonstrate \cite{osborn,kovacs} that 
all eigenmodes of 4d Euclidean Dirac operator in the confined phase  
are delocalized, and hence it behaves as 4d metal.
The metallic property of the Dirac operator in confined phase of QCD is quite counterintuitive
since quark is assumed to be confined in 3d space. 
The only supporting argument of Parisi \cite{parisi} claims that 
the eigenvalues of the Dirac operator in confined phase  have to interact to provide 
the finite density at the origin. Hence they obey the Wigner-Dyson statistics
and therefore the eigenfunctions are delocalized. Let us emphasize once again
that we mean transport not in the physical time but in RG radial time.

We suggest the qualitative explanation of the delocalization in confined phase in terms of the 
topological defects. The starting remark concerns the
identification of the Dirac operator eigenvalue with the imaginary quark mass.
Due to the axial anomaly the mass dependence goes through $m^{N_f}e^{i\theta}$ 
hence to get imaginary masses as required for eigenvalues of the Dirac operator we 
can introduce the  value of $\theta$ which depends on the number
of flavors. In particular for $N_f=0$, when the Dirac operator plays 
the role of probe, $\theta=\pi$ while generically $\displaystyle\theta=\frac{\pi (N_f+2)}{2}$. 

It is known for a while that at $\theta=\pi$ there are  degenerate vacua; hence,
the theory enjoys the domain walls. Recently it was argued
that  theory on the domain wall is in the deconfined phase \cite{zohar}. For instance, if $N_f=0$ the spectral correlator
of the probe Dirac operator tells how quark
could propagate along the domain wall in the pure YM theory
in the radial time evolution. 

This picture qualitatively agrees with the relatively old lattice studies \cite{qcdOLD}
where it was found that the delocalized chiral modes of the Dirac operator 
in confined phase are suited at the $2+1$
surfaces in the 4d Euclidean space which have some rigid 
topological properties. Note that in this scenario we have a kind of fracton
picture (see \cite{fractons} for review). Indeed we have the restricted mobility of 
the percolation type for elementary degrees of freedom in 4d space while  
the composite particles - mesons can propagate freely.

On the top of this conjectured mechanism of  eigenmode  delocalization on "domain wall", there can be additional
mechanism due to the topological delocalization similar to the phenomena
found in \cite{khmel,pruisken,furusaki,kamenev} for low dimensional cases. 
It would be a kind of $4+1$  version of the quantum Hall effect. However,
such additional mechanism potentially can explain only the delocalization  of a single 
mode and can not explain the delocalization of the all modes
of the Dirac operator in the confined phase.

\subsection{Worldsheet arguments}

\begin{figure}
  \includegraphics[width=0.8\textwidth]{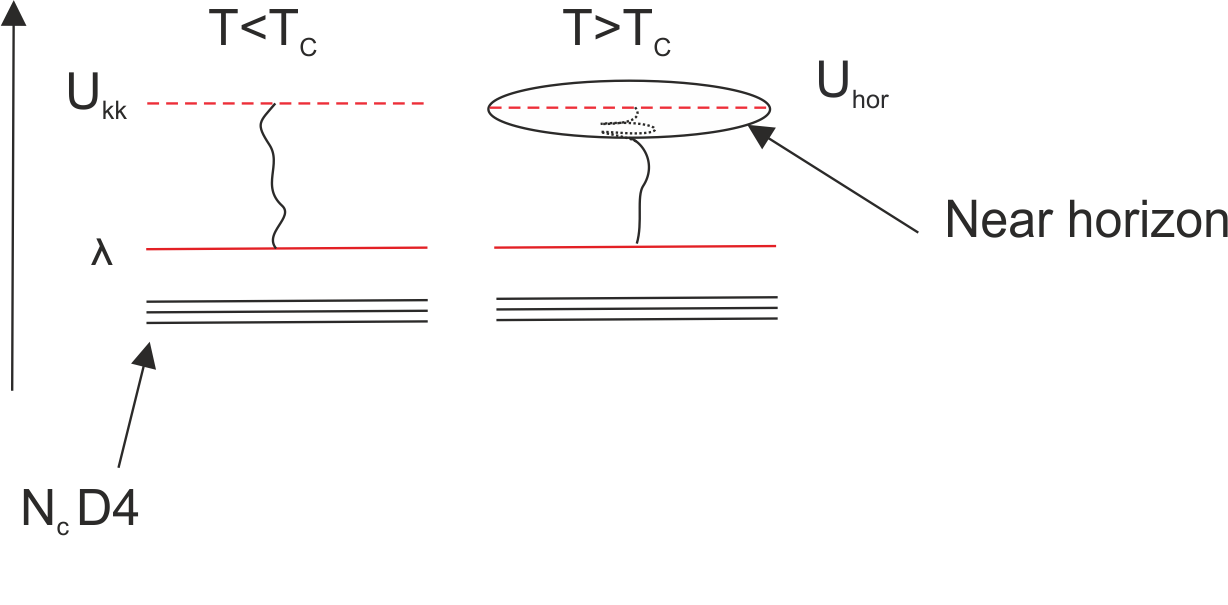}
  \caption{Embedding of the string}
  \label{fig:strbulk}
\end{figure}

In confined phase quark is represented by the  fundamental string extended from the probe flavor
brane located at radial coordinate fixed by $\lambda$ till the effective  IR wall (Fig.\ref{fig:strbulk})
that is worldsheet coordinates are $(r,t_E)$. The relevant target space geometry
involves the cygar in $(r,\phi)$ coordinated and the thermal circle.

Let us first explain the 
holographic interpretation of the Dirac operator eigenvalue. It was 
argued in \cite{stern} that the Euclidean 4D Dirac operator serves
as the Hamiltonian with respect to the Schwinger proper time. On the
other hand, it was shown in \cite{gopakumar,gl} in certain situations that the Schwinger proper
time $ \tau$ is related with the radial coordinate $z$ as $\tau \propto z^2$ in the 
Poincare coordinates 
\beq
ds^2= \frac{dx_idx^i + dz^2}{z^2}
\eeq
for the Euclidean $AdS_{d+1}$ space.
This relation holds in the pure boundary conformal theory \cite{gopakumar} and in the non-SUSY 
boundary theory in the constant external field \cite{gl}. Hence we suggest
that the Dirac operator eigenvalue is conjugated to the radial
coordinate in the AdS-like geometry. This is also consistent 
with the well known identification of the radial position of the
flavor brane corresponding to the quark with mass.

How the spectral statistics in the boundary theory gets translated into the worldsheet framework?
Using the conjectured Dirac operator
spectrum - radial worldsheet coordinate correspondence 
the question of the localization/delocalization  of the 
energy spectrum gets reformulated as the question if 
neighbor  "string bits" are correlated(Wigner-Dyson) or independent(Poisson). In the confined
phase the string has  finite effective tension at all
values of radial coordinate 
\beq
T_{eff}(r) = T_0 \sqrt{g_{tt}(r)g_{xx}(r)}
\eeq
therefore  tensionful string does not have any reason for 
uncorrelated neighbor bits. It is qualitatively consistent with the
Wigner-Dyson  level statistics observed in lattice QCD in confined phase. 
According to the Wigner-Dyson statistics the neighbor string bits are 
repulsive implying that the string tends to be straight.

We conjecture that using the "radial coordinate - Dirac operator eigenvalue"
correspondence
the spectral correlator in the boundary theory gets mapped into correlator 
of the collective fields $\rho(x)$ on the string worldsheet 
\beq
\Braket{\rho(\lambda)\rho(\lambda')}_{boundary}\rightarrow \Braket{\rho(x)\rho(x')}_{worldsheet}
\label{mapping}
\eeq

More arguments supporting this conjecture read as follows

\begin{itemize}

\item The mapping (\ref{mapping}) is known for a while. It was 
demonstrated in  \cite{simons} that the spectral correlator in RMT 
is equivalent to the correlator of the densities in the 2d collective field theory
for the Calogero model describing the fermions in the harmonic 
confining potential \cite{simons}. The   Hamiltonian of the model reads as 
\beq
H_{Cal} =  \sum_{i=1}^{N} p_i^2 + \frac{\beta}{2}\left(\frac{\beta}{2} -1\right)\sum_{i<j}\frac{1}{(x_i - x_j)^2} +\omega^2 \sum_{i=1}^{N}x_i^2
\eeq
and $\rho(x,\tau)$ 
is described in terms of the two-dimensional scalar field as follows \cite{kravtsovrev}
\beq
\rho(x,\tau)= \rho_0 + A\cos (2\pi x + 2\Phi(x,\tau))
\eeq 
Hence the density-density correlator can be expressed in terms of the conventional 
Green function of the scalar $\Phi(x,\tau)$ in $(1+1)$.

\item 

We suggest that collective field $\Phi(s,\tau)$ 
in the worldsheet theory with potential $\cos\Phi$ term appears 
similarly to the case of the string
in the cigar background \cite{kostov} from the condensation of vortices 
near the tip of the cygar. 
The vortices are due to the non-singlet
sector in the $c=1$ matrix model \cite{klebanov,boulatov,maldacenanon}.
It is these vortices or more generically holes in the worldsheet  which 
presumably yield the Calogero 
model and the Luttinger
effective collective field description on the string worldsheet. 

\item

To some extent, we are
looking for the correlator of two Wilson loops. The 
relation between the correlators of the Wilson loops 
and the spectral formfactor has been noted in \cite{sonner} in the context of the matrix model.
In the confined phase the RMT works well hence the interpretation
of spectral correlator
in terms of two Wilson loops or their Laplace transforms - resolvents
is natural. Two Wilson loops correspond to two boundaries of the worldsheet
which could be boundaries of holes or boundary of one hole and the external boundary.
Potentially, we could wonder about the Gross-Ooguri
phase transition \cite{gross} when the connected surface with
two Wilson loop boundaries no longer exists.

\item The spectral statistics is most transparent in terms of the
level spacing distribution. The duality we are looking at implies the identification
of the distribution of the distance between the neighbor levels $P(s)$ and the 
distribution of "`sizes of the
holes"' on the worldsheet $P(\delta z)$. We know that in the delocalized phase 
$P(s)\propto e^{-s^2}$ hence we wonder if $P(\delta z)$ has the similar behavior.
Remarkably it turns out that is the leading approximation the distribution 
of the hole sizes obey this law indeed \cite{abanov}.

\end{itemize}

Let us make a few comments on the spectral density itself. In the 
confined phase we have to explain the Casher-Banks relation from the
worldsheet viewpoint. This has been done to some extend in the 
\cite{kiritsis,sonnen} where the quark condensate was related to the 
tachyon field in the spectrum of open string connecting D8
brane and antibrane. The $\rho(0)$ 
emerges from the tachyon mode of the short open string placed
near the tip of the cigar. This viewpoint is useful for 
identification position of $\lambda=0$ value as the tip of the cigar. 
The linear term in the spectral density is proportional
to the $|\lambda|$ (\ref{density}) which presumably
implies the cusp
in the string shape; however, the special
analysis of this non-analyticity is certainly required.
Moreover, the coefficient in front of the linear term is proportional to the  two-point
correlator of the scalar currents \cite{smilga} which hopefully could help in
a holographic explanation of the non-analyticity of the spectral
density.

\section{Spectral statistics in the deconfined phase}

\subsection{Field theory arguments}

The key question in the deconfined phase concerns 
the holographic interpretation of the mobility edge in the spectrum
found in the lattice studies \cite{osborn}.
Let us present a few qualitative arguments supporting the identification
of the near-horizon region in the holographic dual and the 
localized modes in the Dirac operator spectrum. 
\begin{itemize}
\item
At the critical metal-insulator transition in the Dirac operator spectrum
one could expect the jump of the chiral conductivity in the thermal QCD.
The corresponding Kubo-like formula for the correlator of the 
Noether currents generated left and right chiral rotations reads as 
\beq
i\int dx \Braket{J^{L}_{\nu}(x)J^{R}_{\mu}(0)}  = -\frac{1}{4}\eta_{\mu\nu} F_{\pi}^2
\eeq
Comparison of this QCD low-energy theorem with the 
formulas known in the transport phenomena provides the identification of the $F_{\pi}$ as the
diffusion coefficient in the chiral matter \cite{zahed}. Hence we could question when
the jump of chiral conductivity is expected in the holographic 
setup. To be as model independent as possible consider the anomaly matching 
Son-Yamamoto condition in holographic QCD \cite{son} which yields the relation 
between the 2- and 3-point functions 
and is diagonal with respect to the holographic RG flows in the hard wall model \cite{dgm}.
The Son-Yamamoto relation amounts to the following expression for the "running"$F_{\pi}$
\beq
F_{\pi}^{-2}(z) = \int_0^{z} \frac{1}{f^2(r)}dr,
\eeq
which is valid for any reasonable holographic metric.
\beq
ds^2=-f(z)dt^2+\frac{dz^2}{f(z)}+z^2d\Omega^2.
\eeq
Taking the derivative of this expression we see immediately that
the criticality for the conductivity takes place exactly at the BH
horizon when $f(z)=0$.

\item

The lattice QCD studies demonstrate that the position of mobility edge $\lambda_m(T)$
at $T>T_c$ grows as the function of the temperature near the deconfinement 
phase transition approximately as \cite{temperature}
\beq
\lambda_m(T) = a (T-T_c)
\eeq
with come constant $a$. The possible holographic explanation of this behavior 
goes as follows.
For the near-extremal BH the following  relation 
between the radial coordinate and the temperature
\beq
T \propto (r-r_0),
\eeq
where $r_0$ is the radius of the extremal BH.
This behavior is qualitatively consistent with 
the observed linear temperature dependence of the mobility edge if we assume that
$\lambda \propto (r-r_{kk})$. 
Let us emphasize that the precise holographic metric in the deconfined
QCD is unknown; however, the presence of the horizon is well established. The
detailed lattice study of the dependence $\lambda_m(T)$ could provide some
information concerning the holographic metric.  

\item At the transition point the cigar involving the Euclidean time $t_E$ 
emerges in the bulk which means that  effectively the time circle
is shrinkable and  the theory behaves as 3d theory in
the deconfinement phase. This fits with the fact that  the fractal 
dimension obtained near the mobility edge $\nu= 1.46$ in lattice QCD
corresponds to the 3d unitary Anderson model. On the other hand, the Polyakov loops
which wrap the thermal circle get condensed  and presumably can serve as the 
source of disorder \cite{kovacs}. 

\item
Recently, the interesting relation for the spectral correlator has been 
found \cite{kanazawa}. The analogue of the Casher-Banks relation
for the spectral formfactor reads as follows
\beq
R(\lambda_1=0,\lambda_2=0) -R_{Pois}(\lambda_1=0,\lambda_2=0)= f_A,
\label{kana}
\eeq
where $R(\lambda_1,\lambda_2)$ - is the spectral correlator of the Dirac 
operator in the deconfined phase and $f_A$ is defined as the coefficient 
in front of $U(1)_A$ symmetry breaking term
in the expansion of the partition function in terms of the fermion mass matrix $M$
\beq
Z(M)=\exp\left(-\frac{V_3}{T}\left(f_0 - f_2 \operatorname{Tr} M^{+}M - f_A(\det M + \det M^{+}) +O(M^4)\right)\right)
\eeq
The (\ref{kana})
measures the difference between the  spectral correlator in QCD
and spectral correlator in the case of Poisson statistics. We know
from the lattice studies that near the $\lambda=0$ the statistics
is Poissonian hence this observation implies that $f_A=0$ and unbroken $U(1)_A$ symmetry.
This issue is quite controversial and there are lattice results contradicting
and supporting this statement in the literature. This point certainly
deserves further study.

\end{itemize}

\subsection{Mobility edge and BH near-horizon region. Worldsheet perspective}

Turn now to the identification of the mobility edge in terms 
of the worldsheet theory of the string embedded into the
target space involving BH. 
The bulk BH metrics induces the  thermal metric on the worldsheet. The target and 
worldsheet temperatures coincide for the static quark.
Hence,  we could question what is the natural scale in the worldsheet theory
which separates two parts of the string worldsheet? 
We suggested above that the mobility edge lies in a near-horizon region.

Worldsheet arguments go as follows,
\begin{itemize} 

\item Far from the horizon the string tension is finite and the neighbor elementary string bits interact repulsively yielding
Wigner-Dyson statistics while 
in the near-horizon region the effective tension vanishes and the elementary string bits obey the
Poisson statistics. Similar to confined phase we have to identify the Poisson statistics
of the energy levels with the distribution of the hole sizes. When the tension tends to zero
there are no preferable sizes of holes on the worldsheet hence they indeed enjoy the Poisson
statistics.

\item 
According to our conjecture exactly at
the deconfinement transition temperature $T=T_c$   the critical statistics of the levels in the
boundary theory is expected and is the same at all energies being a mixture
of the Wigner-Dyson and Poisson statistics \cite{shklov}. What does it mean 
for the string worldsheet picture? First, it implies that the level spacing distribution
does not depend on the radial worldsheet coordinate and therefore is RG invariant.
Secondly, the radial distance between two neighbor levels at the boundary corresponds
to the distance between  two neighbor string bits on the worldsheet. The distance 
is measured with the worldsheet metric; hence, naively we could assume that
the worldsheet metric has very peculiar form exactly at the transition point.
There is the transition from the Wigner-Dyson to Poisson at some value of 
the level spacing $s_0$. This means that the finite localization length proportional
to $s_0$ is expected at string worldsheet at the critical point.

\item 
Contrary to the
confined phase the cygar geometry involves now $(r,t_E)$ coordinates and the string extended 
along the radial coordinate evolves in the angular coordinate $t_E$. Thus, 
the thermal effective field theory for Calogero system is the proper starting point. Remarkably, it turns out \cite{ggverba2003} that the RMT-Calogero correspondence at zero temperature
gets generalized to the relation between
the Calogero model at finite temperature and the critical
matrix model \cite{crit2}. Now the
spectral correlator in the critical RMT gets mapped into the density-density correlator
in the Calogero model at finite temperature upon the identification deformation parameter $b$
in the matrix model as  \cite{ggverba2003}
\beq
2b=\frac{\omega}{\sinh(\frac{\omega}{T})} \qquad \text{and} \qquad 2b+1=\frac{\omega \cosh(\frac{\omega}{T}) }{\sinh(\frac{\omega}{T})} 
\eeq
Let us remind that coordinates of particles $x_i$ in Calogero model are identified as radial holographic coordinates while
$\tau$ is the Euclidean time identified with the angular coordinate at the hyperbolic plane.
The density-density correlator is taken at the same Euclidean time. The Calogero model is 
considered at the fixed coupling constant which corresponds to the fermion statistics.

\item
It was observed 
in \cite{kravtsovbh} that the 2d collective field theory yielding the proper criticality 
for the density-density correlators can be curiously identified
with the effective 2d "acoustic BH". 
The fractal dimension $D_2$ derived in
the worldsheet theory is induced
from the bulk involving BH \cite{kravtsovbh} and is related 
to the Hawking  effective temperature T as
\beq
T\propto\pi \frac{d-D_2}{d}
\eeq   

\end{itemize}

\section{Conclusion}

In this Letter we have presented few conjectures concerning the localization 
properties of the Euclidean 4D Dirac
operator spectrum in QCD  in confined and deconfined phases. We suggest that the delocalization
of all modes in the confined phase is related with the $\theta=\pi$- like phenomena.
The quarks are delocalized on the domain walls and can
propagate along them if we treat radial coordinate as time. This fits with the lattice observations 
concerning the location
of delocalized modes on the $2+1$ submanifolds in the Euclidean 4D space \cite{qcdOLD}. 
It was conjectured that in the deconfined phase the localized modes correspond 
to the BH near-horizon region
in the holographic dual. Almost all of our arguments are qualitative  hence further 
quantitative clarification is certainly required.

Our conjecture can be compared with the example of the emergent mobility edge
in the SYK model perturbed by quadratic term which has been found recently in \cite {garcia}. 
It is believed that 2d BH in the dilaton JT gravity  
is dual  to the boundary SYK model \cite{syk} (see \cite{rosenhaus} for review and references). 
The AdS$_2$ geometry has two boundaries and the perturbing term corresponding to the 
fermion mass in (0+1) boundary theory induces the interaction between boundaries
hosting the left and right sectors.
If the strength of interaction is strong enough the mobility edge in the spectrum gets emerged \cite {garcia}.
This phenomena has many similarities with our study in (4+1) since the eigenvalue of the Dirac operator
corresponds to the imaginary mass term, therefore it involves the interaction between 
left and right sectors. If the eigenvalue is large enough the mobility edge can
be identified.

It would be also interesting to match our study with the discussion in \cite{Lee}. It was argued 
there that  quite generically criticality in the ungapped phase of boundary Euclidean theory appears
at the holographic horizon. Its origin is the lack of possibility to match the initial UV state
with the particular state in IR  through the holographic RG flow. 
Our study suggests that the corresponding criticality
at horizon is expected to be the metal-insulator type transition 
in the spectrum of the boundary Euclidean theory.

It would be very 
important to investigate in the lattice QCD framework  the low-energy eigenmodes
of the Dirac operator in the deconfined phase. It would provide the 
important information concerning the near-horizon region of the 
BH in 5D. Another interesting issue concerns the analysis of deconfinement transition
induced by the baryonic density from the viewpoint of Dirac operator spectrum.

\section{Acknowledgments}

We are grateful to  N.Sopenko for collaboration at the early stage of this study  
and to V. Braguta, A. Kamenev, D. Kharzeev, A. Kitaev,
V. Kravtsov, A. Milekhin, S. Nechaev, N. Prokof'ev, B. Shklovskii  for the useful discussions. 
The work of A.G. and M.L. was supported by Basis Foundation Fellowship 17-11-122-21 and RFBR grant 19-02-00214. A.G. thanks SCGP at Stony Brook University and KITP at University of California, Santa Barbara, for the hospitality.

\end{document}